# Coherent Bremsstrahlung Emission in the Early Stage of Relativistic Heavy Ion Collisions

C. Bertulani, L. Mornas, U. Ornik*

Theory Group, GSI, D-64220 Darmstadt, Germany

**Abstract**

We present a model of coherent Bremsstrahlung emission from the compression phase of relativistic heavy ion collisions based on a hydrodynamical description of the collision process. The decceleration of the matter through a shock front, calculated in 1-D relativistic dissipative hydrodynamics, is at the origin of a copious Bremsstrahlung emission. The interferences between the Bremsstrahlung emitted by two receeding shock fronts result in deviations from the usual $1/\omega$ law, with in particular a strong suppression of low energy photons in symmetric collisions. We show how it might be possible to obtain information on the equation of state, possible transition to a quark gluon plasma and dynamical properties of the matter from the study of the interference pattern. We show that the optimal region for the observation of coherent photons is in SIS–AGS energy range.

## 1 Introduction

We would like to present in these proceedings some preliminary results of a macroscopic model of coherent Bremsstrahlung emission in the early stages of relativistic heavy ion collisions. The basic idea of the model is the following: In the compression stage of a heavy ion collision at relativistic energies such as the AGS experiments, two shock waves are expected to develop and propagate through the matter. The deceleration of the matter through the shock fronts can be calculated from a relativistic dissipative hydrodynamical theory [1]. In a symmetrical collision at small impact parameter, the Bremsstrahlung emission coming from the two shock fronts are expected to interfere destructively. This will result in deviations from the commonly accepted $1/\omega$ law of the radiation spectrum. Our model thus differs basically from the standard description [2, 3] where *(i)* "sudden stopping" is assumed, which means that the shock fronts are supposed to be infinitely thin (*i.e.* no dissipation effects), and *(ii)* no interferences are present, since the Bremsstrahlung in this model is arising from the fireball formed immediately after the collision. As a consequence, all the Bremsstrahlung emission in [2] is released suddenly, whereas in our model, the emission extends over a time interval $\tau$ equal to the time necessary for the shock wave to propagate through the nuclear matter.

## 2 The Model

The two basic ingredients of the model are the classical formula of Jackson [4] for Bremsstrahlung emission and the dissipative shock model [1] which describes the deceleration of nuclear matter during a collision. We assume here that the deceleration occurs essentially in the longitudinal (beam) direction, which for high laboratory energies should be a good approximation. The emission spectrum is given in this case by

$$\frac{dI}{d\Omega d\omega} = \frac{Z_t Z_p R^4 \sin^2\theta}{4\omega} \left| \int_0^{t_{max}} dt e^{i\omega t} \int_{-x_{max}}^{x_{max}} dx e^{-i\omega x \cos\theta} \frac{dJ(x,t)}{dt} \right|^2 \quad (1)$$

---

*talk presented at the $5^{th}$ workshop on hadronic interactions, Sao Paulo, 1-3 Dec. 1994

In this formula, $dJ/dt$ is the time derivative of the charge current in the equal velocity system, and is related to the deceleration $dv_\xi/dx$ in the reference system where the shock wave is stationary through a Lorentz boost with the shock velocity $v_{sh}$. $dv_\xi/d\xi$ was obtained earlier [1] from a 1-D relativistic dissipative shock model. In general, when both viscosity and thermal conductivity are taken into account, $dv/d\xi$ is obtained numerically from the step by step integration of a system of differential equations. When thermal conductivity is neglected, we can obtain an analytical solution for the deceleration:

$$\frac{dv_\xi}{d\xi} = \frac{\varepsilon_0 \gamma_{rel}^2 v_{rel}}{4\eta \gamma_\xi^5 v_\xi/3}[(1+c^2)\gamma_\xi^2(1-v_\xi v_{rel}) - 1] \quad (2)$$

Here, $\varepsilon_0$ is the energy density of cold nuclear matter ($\varepsilon_0 \sim 0.2$ GeV.fm$^{-3}$), the propagation velocity of the shock fronts is given by $v_{sh} = c^2/v_{eq}$, with $c^2 = P/\varepsilon$, $v_{rel} = (v_{eq} + v_{sh})/(1 + v_{eq}vsh)$, $v_{eq}$ is the velocity of the colliding nuclei in the equal velocity system, $Z_t$ and $Z_p$ are the charge numbers of the target and projectile nuclei respectively, and $\theta$ is the photon emission angle. The expression is integrated on the space and time intervals defining the extension and duration of the shock (*cf.* [1]):

$$x_{max}(t) = 2 R/\gamma_{eq} - v_{eq} t \quad ; \quad t_{max} = \frac{2 R/\gamma_{eq}}{v_{sh} + v_{eq}} \quad (3)$$

In this short contribution, we limit ourselves to symmetrical collisions of two identical nuclei (for example Au+Au) and small impact parameters ($b \sim 0$), since the effect of interferences which we would like to study here is expected to be maximum in this configuration. Otherwise, the integral has to be carried out on a more complicated space-time element. R is the radius of the target (equal here to the radius of the projectile). As to the choice of the equation of state (eos), we considered four cases: an eos with constant sound velocity $P = \varepsilon/3$, the non-linear QHD-1 model with exchange of $\sigma$ and $\omega$ mesons, a phenomenological fit of the results of lattice QCD simulations [6], and a resonance gas eos including all particles and resonances of the Particle Data Book (up to 2 GeV). The viscosity was kept constant (the currently accepted value is about $\eta \sim 0.05$ GeV/fm$^2$c). For further details of the model, we refer the reader to [1].

## 3  Interference pattern

It is seen from (1) that the overall behavior of the Bremsstrahlung emission spectrum goes as the inverse of the emitted photon energy $1/\omega$. As our purpose is to study the deviations from this law, we will rather use in the following, except when explicitely stated, the dimensionless quantity $\omega d^2 I/(d\omega d\Omega)$. In order to make the model more easily understandable, and study the origin of the interference pattern, we will first use a simplified parametrisation of the derivative of the particle current

$$dJ/dt \propto \delta(x - v_{sh}t) - \delta(x + v_{sh}t) \quad (4)$$

This ansatz, which describes the propagation of two infinitely thin shock fronts, leads to an analytical expression for the Bremsstrahlung spectrum:

$$\omega \frac{dI}{d\omega d\Omega} \propto \text{(target)} + \text{(proj.)} + \text{(interf.)} \tag{5}$$

$$\text{(target, proj.)} = 2\sin^2\theta \frac{(1 - \cos[\omega t_{max}(1 \pm v_{sh}\cos\theta)])}{\omega^2(1 \pm v_{sh}\cos\theta)^2} \tag{6}$$

$$\text{(interf.)} = -2\sin^2\theta \frac{\cos[2\omega t_{max}v_{sh}\cos\theta] - 2\cos[\omega t_{max}v_{sh}\cos\theta]\cos[\omega t_{max}] + 1}{\omega^2(1 - v_{sh}^2\cos^2\theta)}$$

where we have in (6) a $+(-)$ for the projectile and target contribution respectively. In a more realistic scenario the shock fronts are broadened by the dissipative effects (*cf.* §4). In Fig. 1, we display some typical interference patterns with target, projectile and interference contributions for the $P = \varepsilon/3$ equation of state and laboratory energy $E_{lab}$=11.5 GeV. From this picture, we see that the soft part of the spectrum is strongly suppressed by the interference effects. Actually, we have

$$\lim_{\omega \to 0} \omega \frac{d^2I}{d\omega d\Omega} \sim \cos^2\theta \sin^2\theta v_{sh}^2 t_{max}^4 \omega^2 \tag{7}$$

Moreover, it is observed that $dI/(d\omega d\Omega)$ decreases faster than $1/\omega$ for large $\omega$. In general, (5) describes a rich interference pattern with several periods of similar order of magnitudes. Two interesting features however can be extracted from this form.

• If $v_{sh}\cos\theta$ is small (in practice, smaller than $\sim 0.2$), then the spectrum is well reproduced by the fitting formula

$$\omega \frac{d^2I}{d\omega d\Omega} \sim \frac{4\sin^2\theta}{\omega^2} \frac{v_{sh}^2 \cos^2\theta}{[1 - v_{sh}^2\cos^2\theta]^2} \{(\omega t_{max}^2 - 2\omega t_{max}\sin(\omega t_{max}) + 2 - 2\cos(\omega t_{max})\}$$
$$\times \left\{\frac{\sin(\omega t_{max}v_{sh}\cos\theta)}{\omega t_{max}v_{sh}\cos\theta}\right\}^2 \tag{8}$$

which represents a faster oscillating function of period $2\pi/t_{max}$ with an enveloppe ($\sim \sin x/x$) with a larger period $2\pi/(t_{max}v_{sh}\cos\theta)$. (*cf.* Fig. 2a)

• There exists a set of "magical numbers" given by

$$v_{sh}\cos\theta = (n-1)/(n+1); \quad n = \text{integer} \geq 2 \tag{9}$$

The first are 1/3, 1/2, 3/5, 2/3 .... For these values, the minima are very well marked (*cf.* Fig. 2b)

$$\omega \frac{d^2I}{d\omega d\Omega} = 0 \text{ at } \omega = \frac{2\pi n}{t_{max}(1 - v_{sh}\cos\theta)} \tag{10}$$

From the observation of one or several of these features, one could easily deduce the parameters $t_{max}$ and $v_{sh}$ and therefore also $c^2 = P/\varepsilon$ which characterizes the stiffness of the eos.

## 4  Dependence on the equation of state and viscosity

Now we turn to the numerical integration of the full model, with the derivative of the particle current defined from the relativistic dissipative shock model. Figs. 3 and 4 show some typical Bremsstrahlung spectra and their dependence on the parameters of the model, *i.e.*

the laboratory energy and emission angle of the photons, which are known parameters, and the equation of state and viscosity which we are looking for. All results are presented in the equal velocity frame. The general shape of the deviation of spectrum from the $1/\omega$ law consists of one main peak, which location shifts to higher photon energies with higher beam energies, followed by other more or less distinguishable weaker peaks. From Figs. 3a and 3b, we see that the intensity of the emission is strongly dependent on the choice of the equation of state. The emission is also dependent on the viscosity and increases with $\eta$ (*cf.* Fig 4). Additionally we examined the integrated intensities $dI/d\Omega$, $dI/d\omega$ and $I$. Preliminary calculations indicate that the integrated intensity increases with the shock velocity and the duration of the shock according to a very simple formula:

$$I \propto t_{max}^2 v_{sh}^{1.63} \tag{11}$$

The strong dependance of the photon intensity on the stiffness of the eos makes it specially interesting to investigate the possible occurence and critical energy density and temperature of a phase transition. We have seen in [1] that a sharp decrease in $c^2$ from the lattice eos corresponding to the phase transition was occuring for beam energies $\sim 5$ GeV. Since $t_{max}$ reaches its highest values in the same beam energy range $E_{lab} \sim 5$ GeV [1] and the intensity is roughly proportional to $t_{max}^2$, the conditions of observation will be optimal. Taking into account the uncertainties in the determination of the eos, we propose to measure Bremsstrahlung in the SIS ($\sim 2$ AGeV) - AGS ($\sim 15$ AGeV) energy range.

## 5  Conclusion

In this contribution, we have presented some preliminary results on the coherent Bremsstrahlung emission from the early (shock) stages of a heavy ion collision. The stopping of the matter was obtained from a relativistic a dissipative shock model. Thus, our model does not make the "sudden stopping" hypothesis of Kapusta but allows the description of interferences. The study of the deviations of the spectrum from the standard $1/\omega$ law indicates several ways through which one can determine the equation of state and viscosity of the matter. A measurement of soft Bremsstrahlung-photons in the SIS-AGS energy range could therefore provide with direct informations about the nuclear dynamics.

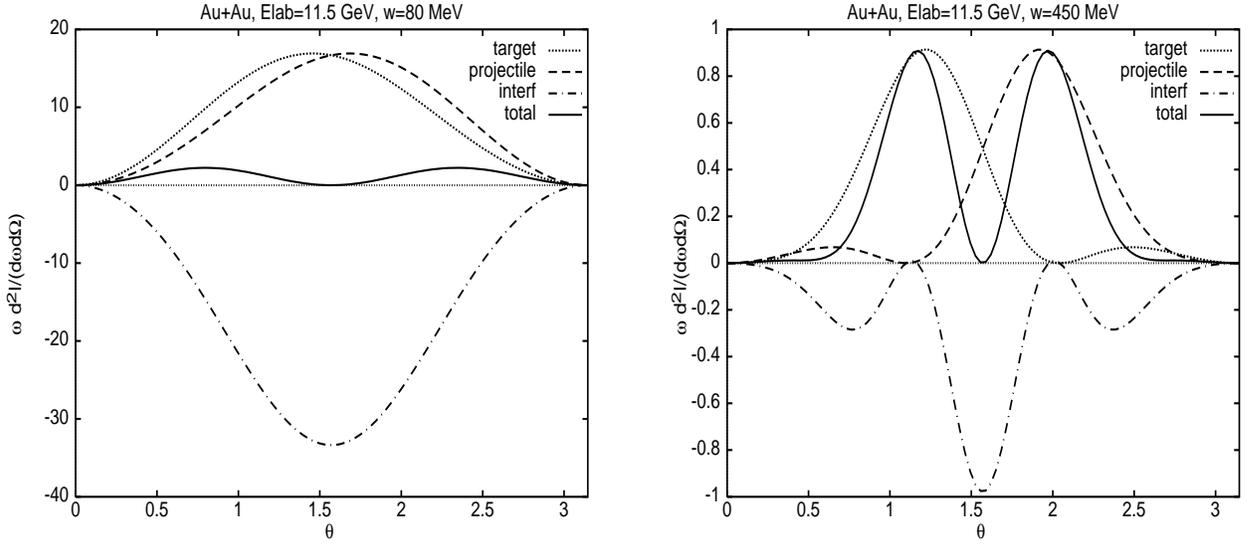

Figure 1: Photon intensity contribution from target, projectile and interference as a function of the emission angle for two photon energies $E_\gamma$=80 and 450 MeV

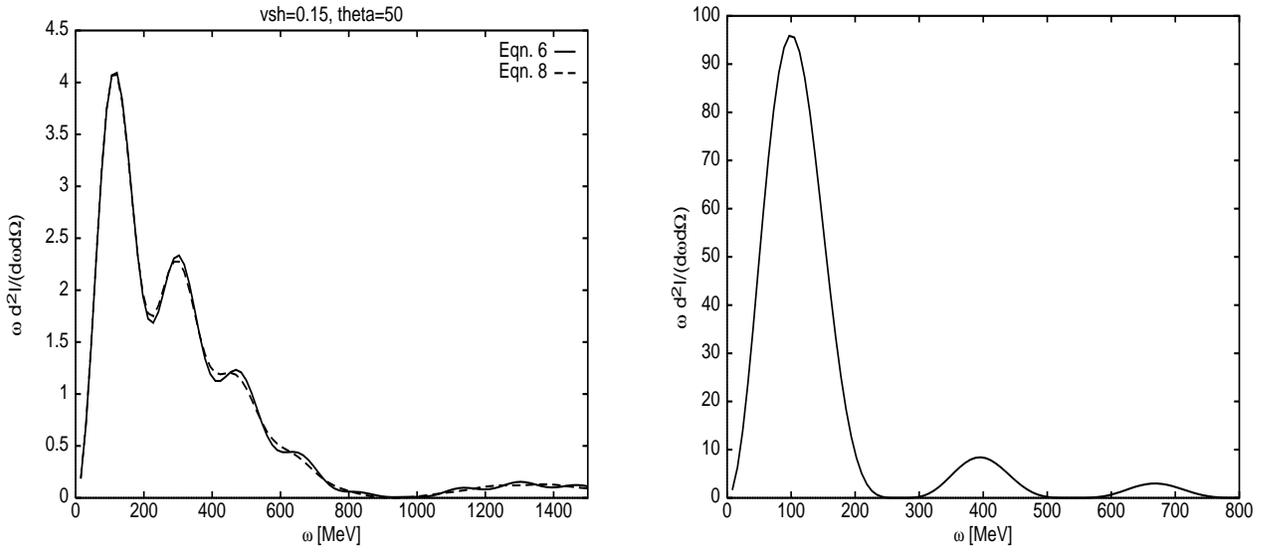

Figure 2: Model interference spectrum (a) Fitting formula for $v_{sh}\cos\theta \leq 0.2$ (b) extinction for "magic" number $n = 2, v_{sh} = 0.47, \theta = 45^o$

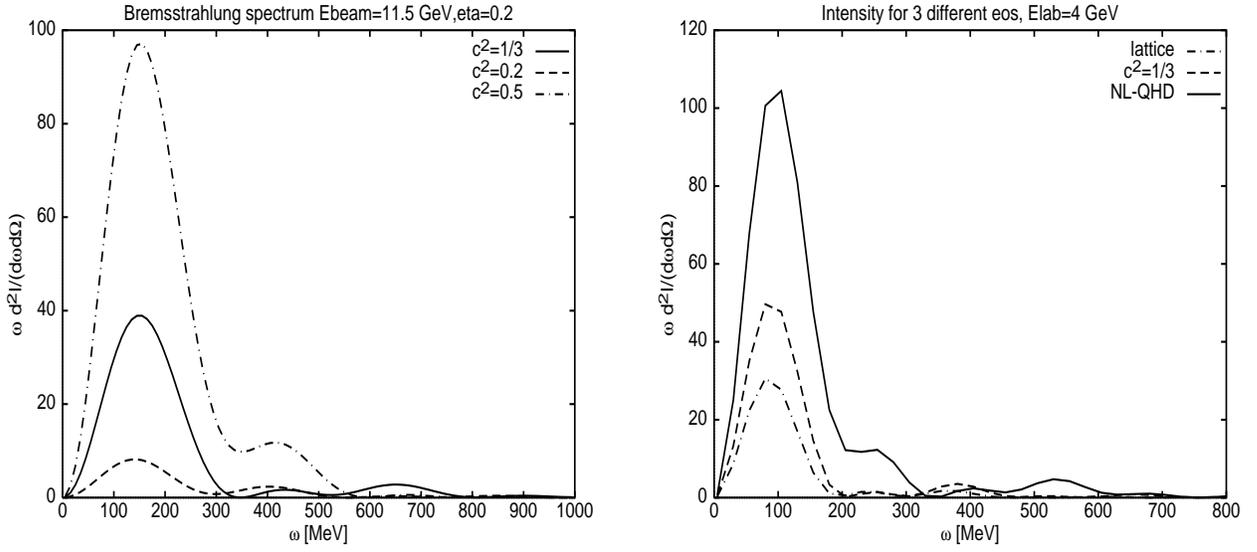

Figure 3: Photon intensity for various choices of the eos

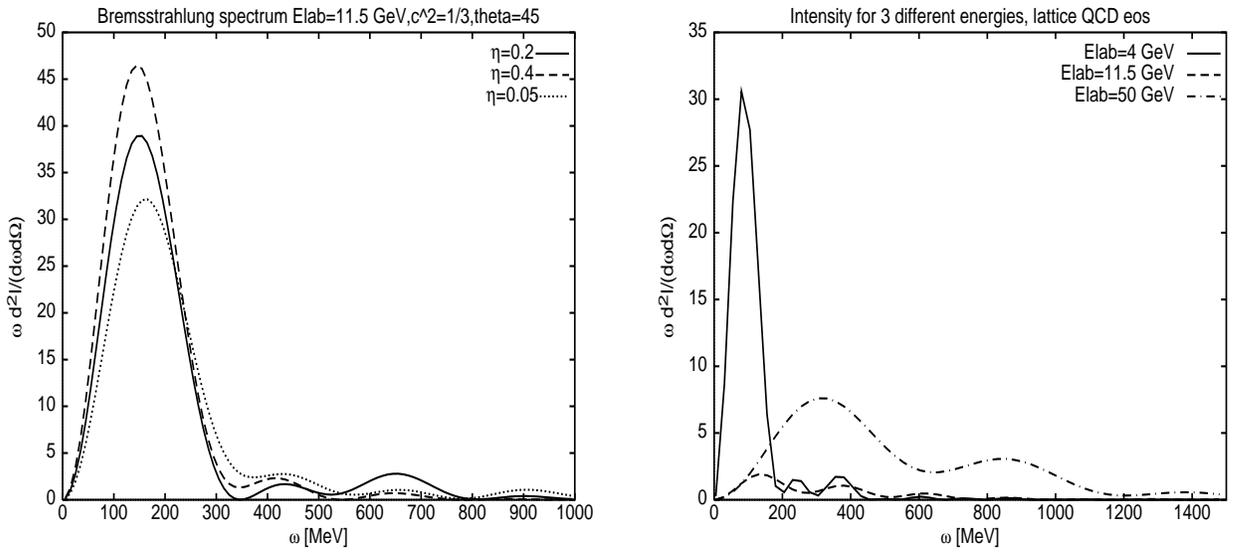

Figure 4: Photon intensity for various values of the viscosity (a) and beam energy (b)